\documentclass{ws-p8-50x6-00}

\begin{document}

\title{R-parity Violation and Neutrino Masses}

\author{Eung Jin Chun\footnote{Presented at COSMO99, Trieste, Italy}}

\address{ Korea Institute for Advanced Study, 
    Seoul 130-012, Korea} 


\maketitle

\abstracts{
R-parity violation in the supersymmetric standard model 
could be the origin of neutrino masses and mixing accounting for
the atmospheric and solar neutrino oscillations.
More interestingly, this hypothesis may be tested in future colliders
by detecting lepton number violating decays of the lightest
supersymmetric particle.  Here, we present a 
comprehensive analysis for the determination of sneutrino 
vacuum expectation values from the one-loop effective scalar potential, 
and also for the one-loop renormalization of neutrino masses and mixing.  
Applying our results to theories with gauge mediated supersymmetry 
breaking, we discuss the effects of the one-loop corrections and
how the realistic neutrino mass matrices arise.
}

The minimal supersymmetric standard model (MSSM) may allow for 
explicit lepton number and thus R-parity  violation through which
neutrinos get nonzero masses \cite{HS}.  As it is an attractive 
possibility to explain the neutrino mass matrix
consistent with the current data coming  from, in particular,
the atmospheric \cite{skam} and solar neutrino \cite{solan} experiments, 
many works have been devoted to investigating the properties of neutrino 
masses and mixing arising from R-parity violation 
[see references in \cite{kang}].
The lepton number and R-parity violating terms 
in the MSSM superpotential  are
\begin{equation} \label{supo}
 W=\mu_i L_i H_2 + \lambda_{ijk}L_i L_j E^c_k + 
     \lambda'_{ijk} L_i Q_j D^c_k \,.
\end{equation}
We first recall that there are two contributions to  
neutrino masses from R-parity violation.  One is the loop
mass arising from one-loop diagrams with the exchange of 
squarks, sleptons or gauginos.  
The other is the tree mass arising from the misalignment of 
the bilinear couplings in the superpotential and 
sneutrino vacuum expectation values (VEVs)
determined by the minimization of the
scalar potential,
\begin{equation} \label{scpo}
  V= [ (m^2_{L_i H_1}+\mu\mu_i) L_i H_1^\dagger + 
       B_i L_i H_2 + {\rm h.c.} ] + m^2_{L_i} 
      |L_i|^2 + V_D + \cdots \,.
\end{equation}
Here $m^2_{L_i H_1}$, $B_i$ and $m_{L_i}^2$ are the soft terms, 
$\mu$ is the supersymmetric Higgs mass parameter, and 
$V_D$ denotes the SU(2)xU(1) D-terms.

The first attempt to investigate whether R-parity violation 
can provide the solution to the atmospheric and solar neutrino 
problems has been made in \cite{hemp}, 
where it was found that the minimal supergravity model
with bilinear R-parity violating terms naturally yields the 
desired neutrino masses and mixing angles.   According to the scatter plot 
study of minimal supergravity parameter space \cite{hemp}, 
the matter conversion (vacuum oscillation) solution to the solar 
neutrino problem was found to be realized in a few \% (20 \%) of 
the selected parameter space.
After the observation of muon neutrino oscillation in Super-Kamiokande 
\cite{skam},  the similar attempt has been made in the context of
minimal supergravity models with generic (trilinear) R-parity violating 
couplings \cite{chun} to find out the preferred ranges of the 
soft parameters depending on $\tan\beta$, and to obtain the correlation
between the neutrino properties (the atmospheric mixing angle and
the ratio between two heaviest masses) and the soft parameters.

A remarkable feature of R-parity violation
as the origin of neutrino masses and mixing is that this idea can
be tested in future collider experiments despite the small R-parity
violating couplings.  For instance, the large mixing angle between
the muon and tau neutrino \cite{skam} implies the observation
of comparable numbers of muons and taus produced in the decays of 
the lightest supersymmetric particle (LSP) \cite{viss}.  Moreover,
the factorization of the R-parity even and odd quantities in 
the neutrino-neutralino mixing matrix enables us to probe
the $\nu_\mu$--$\nu_\tau$ and $\nu_\mu$--$\nu_e$ oscillation amplitudes
measured in Super-Kamiokande \cite{skam} and the CHOOZ experiment 
\cite{chooz} directly in colliders through the measurement of 
the electron, muon and tau branching ratios of the neutralino LSP 
\cite{jaja}.   In a detailed analysis, it was found that 
this testability holds in most of $\tan\beta$ and neutralino 
mass parameter space, in particular,
for the case of the bilinear R-parity violating models \cite{choi}.

\medskip

More recently, further development has been made to include the one-loop 
effect in the determination of the sneutrino VEVs 
\cite{kang,romao}.  For this, one adds the one-loop correction,
\begin{equation} \label{V1}
 V_1={1\over 64\pi^2} 
 {\rm Str}{\cal M}^4 \left( \ln{{\cal M}^2\over Q^2} -{3\over2}\right)\,,
\end{equation}
to the scalar potential (\ref{scpo}).
Then, it is straightforward from the one-loop effective
scalar potential to calculate the sneutrino VEVs,
\begin{equation} \label{svev}
 {\langle L_i^0 \rangle \over \langle H_1^0 \rangle } = 
  -{ B_i \tan\beta + (m^2_{L_i H_1} +\mu\mu_i) 
   + \Sigma^{(1)}_{L_i} \over 
  m^2_{L_i} + {1\over2} M_Z^2 c_{2\beta} + \Sigma^{(2)}_{L_i} }\,,
\end{equation}
where the one-loop correction terms
$\Sigma^{(1,2)}_{L_i}$ are  given by
\begin{equation} \label{Sigme}
\Sigma^{(1)}_{L_i} = 
\left.{\partial V_1 \over H_1^{0*} \partial L_i^0}\right|_{L^0_i=0}, \quad
\Sigma^{(2)}_{L_i} = 
\left.{\partial V_1 \over L_1^{0*} \partial L_i^0}\right|_{L^0_i=0}  \,,
\end{equation}
under the condition that 
the R-parity violating parameters are small, $\mu_i/\mu, \lambda,
\lambda' \ll 1$, which is the case with small neutrino masses, $m_\nu \ll M_Z$.
The essential step in determining $\Sigma^{(1,2)}$'s is the diagonalization of 
the mass matrices of neutralinos/neutrinos, charginos/charged leptons and 
Higgses/sleptons which get mixed due to R-parity violation.  This procedure
can be done analytically when the R-parity violating parameters are small.
The complete analytic formulae for the $\Sigma$'s 
are then obtained in \cite{kang} including the contributions of 
all the particles in the MSSM.  Having determined the sneutrino VEVs, 
one obtains the tree mass given by 
\begin{equation} \label{mtree}
 M^\nu_{ij} = {M_Z^2 \over F_N} \xi_i \xi_j c_\beta^2 
 \quad\mbox{with}\quad
 F_N=-{M_1M_2 \over M_1 c_W^2+M_2s_W^2}- {M_Z^2\over\mu}s_{2\beta}
\end{equation}
where $\xi_i\equiv \langle L^0_i \rangle/\langle H_1^0 \rangle-\mu_i/\mu$,
and $M_{1,2}$ are the masses of the neutral gauginos, bino and wino, 
respectively. Recall that the matrix (\ref{mtree}) makes
only one neutrino massive. 

To obtain the complete neutrino mass eigenvalues and mixing angles,
one needs to perform one-loop renormalization of the neutrino/neutralino
mass matrix through the general formula,
\begin{equation} \label{Mpole}
 M^{pole}(p^2) = M(Q)+ \Pi(p^2) - 
{1\over2}\left( M(Q) \Sigma(p^2)+ \Sigma(p^2) M(Q)\right)
\end{equation}
where  $Q$ is the renormalization scale, and $M(Q)$ is the
the $\overline{DR}$ renormalized tree-level mass matrix, $\Pi$ 
and $\Sigma$ are the contributions from one-loop self-energy diagrams.
In our case, $M$ is the 7x7 neutrino/neutralino mass matrix consisting of
the 3x4 neutrino-neutralino mixing mass matrix $M_D$ and  
the 4x4 neutralino mass matrix $M_N$. 
To find out the 3x3 neutrino mass matrix, usually used is the on-shell
renormalization scheme \cite{hemp,romao}
in which one works in the tree-level mass basis and
rediagonalizes the one-loop corrected 7x7 mass matrix (\ref{Mpole}).
But, as far as only the neutrino sector is concerned, it is more useful 
to work in the weak basis and obtain the effective neutrino mass
matrix by one-step diagonalization \cite{kang}.
Following this procedure, one finds the neutrino mass matrix;
\begin{eqnarray} \label{mpole}
M^{\nu} (p^2) &=&  -M_D M^{-1}_N M_D^T(Q)  + \Pi_n\!(p^2)   \nonumber \\
 &+&M_D M^{-1}_N \Pi_D^T\!(p^2)+ \Pi_D\!(p^2) M^{-1}_N M_D^T
\end{eqnarray}
neglecting the subleading contributions.
Here, $\Pi_{n,D}$  are the neutrino-neutrino and 
neutrino-neutralino one-loop masses.
Note that the first term on the right-hand side of Eq.~(\ref{mpole}) is the 
tree mass given in Eq.~(\ref{mtree}). As can be seen from (\ref{mpole}),
there is no need to calculate the other one-loop terms such as
$\Pi_N$ for neutralino self-energies and  $\Sigma$'s.  Furthermore,
we can simply take $p^2=m_\nu^2=0$ to calculate the physical neutrino
masses and mixing.  This should be contrast with the on-shell renormalization
scheme where the neutralino masses are also involved in calculating the 
one-loop renormalized neutrino masses.

\medskip

Applying the above results to the theories with gauge mediated supersymmetry 
breaking, one can find various interesting properties \cite{kang}.  
First of all,  the one-loop corrections 
$\Sigma_{L_i}^{(1,2)}$ in Eq.~(\ref{svev}) can lead to
a drastic change in the sneutrino VEVs. This effect is magnified in the 
parameter ranges where a partial cancellation between the two quantities, 
$(m^2_{L_iH_1}+\mu\mu_i)$ and $B_i \tan\beta$, occurs.  In certain cases,
it can even change the order of magnitudes of the neutrino mass eigenvalues.
The most interesting question is again what kind of  
realistic neutrino mass matrices accounting for the atmospheric 
and solar neutrino oscillations can be obtained.
In the bilinear models, it turns out that only realizable is 
the small mixing MSW solution to the solar neutrino 
problem with $\tan\beta \geq 20$.
In the trilinear models, one has more freedom and thus more solutions.
For instance, the small mixing MSW solution can be obtained even 
for small $\tan\beta$.  An interesting aspect is that  the bi-maximal mixing 
solution to the atmospheric and solar neutrino problems 
can also be realized with small $\tan\beta$.

Finally, we would like to emphasize that the desirable neutrino 
mass matrices with R-parity violation arise without fine-tuning of 
parameters given the overall smallness of R-parity violating parameters, 
$\mu_i/\mu, \lambda$ and $\lambda'$.
This smallness may be a consequence of a certain flavor symmetry 
responsible for the quark and lepton mass hierarchies, such as
horizontal U(1) symmetry \cite{kchoi}.

\end{document}